
%
%
%
%

\input harvmac

\Title{\vbox{\hbox{CTP \# 2190}\vskip 0.1in\hbox{HUTP-92/A054}}}
{The Hunting of the MR Model}

\centerline{Nuria Rius$^a$\footnote{$^*$}{e-mail: mitlns::rius,
huhepw::simmons} and Elizabeth H. Simmons$^{b *}$}
\bigskip\centerline{$^a$Center for Theoretical Physics}
\centerline{Laboratory for Nuclear Science and Department of
Physics}
\centerline{Massachusetts Institute of Technology}
\centerline{Cambridge, MA 02139}
\bigskip\centerline{$^b$Lyman Laboratory of Physics}
\centerline{Harvard University}
\centerline{Cambridge, MA 02138}

\vskip .3in

We consider experimental signatures of the standard model's minimal
supersymmetric extension with a continuous $U(1)_R$ symmetry (MR model).
We focus on the ability of existing and planned  electron-positron
colliders to probe this model and to distinguish  it from both the standard
model and the standard model's minimal supersymmetric extension with a
discrete $R$-parity.

\Date{3/24/93}

%
\def\phot{$\tilde\gamma$}
\def\hgam{$\tilde H_{\gamma}$}
\def\gae{\raise-.5ex\vbox{\hbox{$\; >\;$}\vskip-2.9ex\hbox{$\; \sim\;$}}}
\def\lae{\raise-.5ex\vbox{\hbox{$\; <\;$}\vskip-2.9ex\hbox{$\; \sim\;$}}}

%
\nref\mrfirst{L.J. Hall and L. Randall, {\sl Nucl. Phys.} {\bf B352} (1991)
289}
\nref\radtop{J. Ellis, G. Ridolfi and F. Zwirner, {\sl Phys. Lett.} {\bf
B257} (1991) 83; H.E. Haber and R. Hempfling, {\sl Phys. Rev. Lett.} {\bf
66} (1991) 1815}
\nref\mrreborn{L. Randall and N. Rius, {\sl Phys. Lett.} {\bf B286} (1992)
299}
\nref\barger{V. Barger, K. Cheung, R.J.N. Phillips and A.L. Stange,
{\sl Phys. Rev.} {\bf D46} (1992) 4914}
\nref\lep{ALEPH Collaboration, {\sl Z. Phys.} {\bf C53} (1992) 1}
\nref\pdg{Particle Data Group, {\sl Phys. Rev.} {\bf D45} No. 11-II
(1992) IX.9}
\nref\aleph{ALEPH Collaboration, {\sl Phys. Lett.} {\bf B265} (1991) 475;
{\sl Physics Reports} {216} (1992) 253}
\nref\lw{B.W. Lee and S. Weinberg, {\sl Phys. Rev. Lett.} {\bf 39} (1977)
165}
\nref\slw{S.L. Wu et al., {\sl ECFA Workshop on LEP 200} CERN 87-08
(1987) 312}
\nref\delphi{S. Katsanevas et al., {\sl Report from the Working Group
on LEP 200 Physics} DELPHI note 92-166 PHYS 250 (December 1992)}
\nref\jangri{P. Janot, Orsay preprint LAL 91-61 (1991); J.-F. Grivaz,
Orsay preprint LAL 91-63 (1991)}
\nref\ehlq{E. Eichten, I. Hinchliffe, K. Lane and C. Quigg, {\sl
Rev. Mod. Phys.} {\bf 56} (1984) 579}
\nref\cdref{C. Dionisi et al. {\sl Aachen ECFA Workshop 1986} (1987) 380}

%
\nfig\zerofigg{Mass of the CP-even neutral Higgs boson as a function
of $\tan\beta$, including one loop radiative corrections with
the photino cosmological constraint incorporated, for
$m_t=$ 140 GeV (solid), 160 GeV (dashed), 180 GeV (dashed-dotted)
and 200 GeV (dotted).
For each value of the top mass, the lower curve corresponds to
$m_A=10$ GeV and the upper curve to $m_A=1$ TeV.}

\nfig\firstfigg{CP-odd Higgs boson partial decay widths
for $m_t = 160$ GeV and $\tan\beta = 0.71$.
The decay modes are
$f \bar{f}$ (A),
$\tilde {H}_\gamma \tilde{z}$ (B),
$\tilde {w}^+ \tilde{w}^-$ (C),
$h Z$ (D) and
$\tilde {H}_Z \tilde{z}$ (E).}

\nfig\secfigg{Excluded regions of the ($m_A, \tan\beta$) plane
inferred from the ALEPH results at LEP I.
Solid curves correspond to $Z \rightarrow Z^* h$ searches
for $m_t=$130 GeV (A), 140 GeV (B), 160 GeV (C) and 180 GeV (D).
Discontinuous curves correspond to
$Z \rightarrow A h$ searches
for $m_t=$ 180 GeV (dashed) and 200 GeV (dashed-dotted).
Areas above or to the left of the curves are excluded.}

\nfig\thirdfigg{Regions in the ($m_A, \tan\beta$) plane accessible to
LEP II, for $m_t=$ 170 GeV (dashed), 180 GeV (dashed-dotted) and
200 GeV (solid).
Areas to the right of the curves are inaccessible.
The dotted curves on the left side of the plot show the region of the
($m_A, \tan\beta$) plane where the difference between the number of
$Z h$ and $Z h_{SM}$ events is larger than $3\sigma$
for $m_t=$ 180 GeV (A) and 200 GeV (B).
In the areas above or to the left of the dotted curves it should be
possible to disentangle the MR $h$ boson from $h_{SM}$.}

\nfig\sixfigg{Same as \thirdfigg with $b$ quark tagging,
assuming $\sqrt{s}= 190$ GeV,
for $m_t=$ 180 GeV (dashed-dotted) and 200 GeV (solid);
and assuming $\sqrt{s}= 200$ GeV for $m_t=$ 200 GeV (dashed).}

\nfig\fourfigg{Contours showing how number of
$Ah \rightarrow \tau \tau j j$
events at NLC depends on $m_A$ and $\tan\beta$.
Curves shown are for $m_t$ = 160 GeV; numbers of
events for each curve are: A=90, B=80, C=70, D=60, E=50, F=40, G=30.  As
$m_t$ is varied, the upper curves simply correspond to an altered
number of events.  For $m_t$ = 180 GeV, one has A=120, B=110, C=100,
D=90, E=80, F=70.  For $m_t$ = 140 GeV, one has C=40, D=30;
in this case, the plane extends
only up to $\tan\beta$ = 1.2, as discussed in Section 3.}

\nfig\fivefigg{Maximum slepton mass (assuming all the sleptons are
degenerate) as a function of $\tan \beta$, for different values of $m_t$:
140 GeV (solid), 160 GeV (dashed), 180 GeV (dashed-dotted) and
200 GeV (dotted). The constraint $m_h \ge 44$ GeV has been
incorporated.}

\newsec{Introduction}

We consider experimental signatures of the standard model's
minimal supersymmetric extension with a continuous  $U(1)_R$ symmetry.
When this class of models was first considered \mrfirst ,
LEP data on the non-observation of light Higgs bosons appeared to exclude
the minimal such model (MR model).
Once the large size of radiative corrections to Higgs masses from
a very heavy top quark became apparent \radtop, the model
was reexamined \mrreborn\ and pronounced viable.

We focus here on the ability of existing and planned electron-positron
colliders to test the Higgs boson and slepton sectors of the MR model.  We
also discuss the extent to which it is possible to experimentally
distinguish the  MR model from both the standard model (SM) and its
minimal supersymmetric extension with a discrete R-parity (MSSM).   As
such, our work was partly inspired by and is complementary to recent papers
on the prospects for studying the Higgs sector of the MSSM at various
colliders \barger.  Our results also complement previous work on MR model
phenomenology.  Experimental signatures involving gaugino production were
considered in \mrfirst ; these are useful because they apply equally well
to many non-minimal $U(1)_R$ symmetric models.   The phenomenological
implications of cosmological constraints on the lightest superpartner (the
photino) were considered in \mrreborn ; those results will be revisited and
applied in this work.

The second section of this paper introduces the model and indicates how
cosmological constraints on the photino mass influence the model's
phenomenology.   Section 3 then discusses the masses and decay modes of the
Higgs bosons  for which we propose to search.   In the next three sections,
we discuss the extent to which current LEP I data and proposed LEP II and
NLC searches for neutral Higgs bosons can and will constrain the MR model.
Slepton searches are the subject of Section 7. Section 8 summarizes our
results and compares the expectations for the MR, SM and MSSM models.

\newsec{The MR Model}

The model we study is the standard model's minimal supersymmetric extension
with a continuous $U(1)_R$ symmetry.  We define the continuous $R$ symmetry
by giving the coordinate of superspace, $\theta$, and all matter
superfields charge +1 while all Higgs superfields have charge 0.
Expansions of the superfields in terms of the component fields then show
that all ordinary particles are $R$ neutral while all superpartners carry
non-zero $R$ charge.  Since the $U(1)_R$ symmetry forbids Majorana gaugino
masses, the model contains an additional field to give a Dirac mass to the
gluino. This field appears only in the soft supersymmetry-breaking sector
and is irrelevant to the rest of this paper; we therefore omit it here and
refer the reader to \mrfirst\ for details.

The most general Lagrangian consistent with the above assumptions is
described by the superpotential:
\eqn\spot{f\ =\ U^c\lambda_U\,Q\,H_2\ +\ D^c\lambda_D\,Q\,H_1\ +\
E^c\lambda_E\,L\,H_1}
where each term has $R=2$ and the quark and lepton superfields $Q, U^c,
D^c, L, E^c$ have the usual $SU(3)\times SU(2)\times U(1)$ gauge
interactions.  The most general soft supersymmetry breaking potential
consistent with our symmetries and a GIM-like mechanism to naturally
suppress flavor-changing neutral currents is:
\eqn\soft{\eqalign{{\cal{L}}_{soft}\ =\ &
m_{H_1}^2 H_1^* H_1 + m_{H_2}^2 H_2^* H_2
+ m_Q^2 \tilde Q^* \tilde Q + m_{U^c}^2 \tilde U^{c*} \tilde U^c
+ m_{D^c}^2 \tilde D^{c*} \tilde D^c\ + \cr
& m_L^2 \tilde L^* \tilde L\ + m_{E^c}^2 \tilde E^{c*} \tilde E^c
+\ B H_1 H_2 + ... \cr}}
where we neglect small Yukawa-suppressed corrections to superpartner
masses.  Notice that we do not assume that
all superpartners have the same mass.

The gaugino mass matrix has a simple form in the MR model
because  \mrfirst\ there is neither a $H_1 H_2$ mass term in the
superpotential (i.e. $\mu = 0$) nor any Majorana gaugino mass in the
supersymmetry breaking potential (i.e. $M, M' = 0$).
The only mass terms are the Dirac masses coupling the partners of the
electroweak gauge bosons to the fermionic partners of the Higgs.
As is conventional, we shall
define a mixing angle $\beta$ in terms of the
ratio of the Higgs VEVs ($v_2/v_1 \equiv \tan\beta$) and shall
denote the superpartner of the $W$ gauge bosons as $\tilde w$ and that of
the $B$ gauge boson as $\tilde b$.
Then the zino ($\tilde z \equiv -
\sin\theta_W \tilde b + \cos\theta_W \tilde w_3$) and one of the
higgsinos
($\tilde H_Z \equiv \cos\theta_W \tilde b + \sin\theta_W \tilde w^3$)
combine to form a Dirac fermion which has
the same mass as the $Z$ gauge boson at tree level, while the photino and
$\tilde H_\gamma$, the higgsino orthogonal to $\tilde H_Z$,
are massless.
The charginos have tree-level masses
$m_{+} = \sqrt{2} m_W \cos\beta$ and  $m_{-} = \sqrt{2} m_W
\sin\beta$.  At one loop, the alteration of the mass structure most
significant for our analysis is the generation of a Dirac mass for the
photino and its associated higgsino, $\tilde H_\gamma$:
\eqn\photmas{m_{\tilde\gamma} =
1.3 {\rm GeV}\ \cot\beta \left({m_t\over 200{\rm GeV}}\right)^2
\left|{m_{\tilde t_L}^2 \over m_{\tilde t_L}^2 - m_t^2} \ln
            {m_{\tilde t_R}^2 \over m_t^2} -
      {m_{\tilde t_R}^2 \over m_{\tilde t_R}^2 - m_t^2} \ln
            {m_{\tilde t_R}^2 \over m_t^2}\right|.}
Note that the photino mass decreases as the top squarks $\tilde t_L$ and
$\tilde t_R$ become more nearly degenerate.

Unlike the MSSM, the MR model is strongly constrained by photino
phenomenology.  To begin with,  the small size of the photino mass
\photmas\ makes  the decay $Z \rightarrow$\phot\hgam\  possible; this, in
turn, renders  the invisible Z width larger than the standard  model value.
The  Z branching fraction into photino plus higgsino will be suppressed by
a factor of $\cos^2 2\beta$ relative to the branching into one SM neutrino
flavor.  Therefore, the bound on the Z width \lep\ $\Delta\Gamma_Z /
\Gamma_\nu < 0.11$ at $2\sigma$ translates into the  constraint $|\cos
2\beta| < 0.33$. This implies that  only the range $0.71 < \tan \beta <
1.41$ is allowed in the MR model.  We will apply this constraint from here
on.

In addition, the cosmological constraint
($\Omega_{\tilde\gamma} h^2 \leq 1$)
on the present photino mass density must be satisfied.
The upper bound on $\Omega_{\tilde\gamma}$ provides a lower bound on the
cross-section for photino annihilation ($\sigma_{\tilde\gamma}$).
Because $\sigma_{\tilde\gamma}$
grows as $m_{\tilde\gamma}^2$, very {\sl light} photinos will yield too
large a residual mass density.
Further, since both s-channel $Z$ exchange and
t-channel slepton exchange are required to make $\sigma_{\tilde\gamma}$
large enough, the sleptons can weigh no more than about 140 GeV
in the MR model (see \fivefigg\ and \mrreborn ); this point will be
crucial to our discussion of slepton searches in Section 7.
To determine the lightest allowed photino mass
(the one giving $\Omega_{\tilde\gamma} h^2 =1$), we maximize the slepton
contribution to $\sigma_{\tilde\gamma}$ by making the sleptons degenerate
at the lightest experimentally \pdg\ allowed mass (65 GeV).
Applying this lower bound on $m_{\tilde\gamma}$ to equation \photmas\
has two important consequences for the MR model.

First, when the top quark is relatively light  ($\lae 150$ GeV),  producing
a heavy enough photino requires $\tan\beta$ to be smaller than some maximum
value, which is obtained when  the top squark masses are as widely
separated as possible (making one mass 1 TeV and the other, $m_t$) so as to
maximize the top squark contribution to \photmas. This constraint is
independent of $m_A$, and it further  restricts the allowed range of
$\tan\beta$. We find $\tan\beta < 0.95 (1.2)$  for $m_t = 120 (140)$ GeV.

The second effect arises more indirectly. Given the form of equation
\photmas , the existence of a lower bound on $m_{\tilde\gamma}$ prevents
the top squarks from being degenerate in the MR model.  We will see in
Section 3 that this reduces the maximum size of the radiative corrections
to the Higgs masses relative to the the maximum size in the MSSM. As we
will discuss in Section 5, the lighter CP-even neutral Higgs boson of the
MR model is therefore accessible to LEP II in a wide region of the
$\tan\beta$ vs $m_A$ plane even if the top quark is very heavy.

\newsec{Neutral Higgs Bosons in the MR Model}

The relations among the tree-level masses of scalar ($H^0_1, H^0_2$),
pseudoscalar ($A^0$) and charged ($H^+, H^-$)
Higgs fields are identical to
those in the MSSM.  In terms of the variables $m_A$ and $\beta$,
the tree-level mass matrix for $H^0_1$ and $H^0_2$ is
\eqn\masss{\pmatrix{M_Z^2 \cos^2\beta + m_A^2 \sin^2\beta&-\sin\beta
\cos\beta (M_Z^2 + m_A^2)\cr -\sin\beta \cos\beta (M_Z^2 + m_A^2)&
M_Z^2 \sin^2\beta + m_A^2 \cos^2\beta\cr}\ \ .}
At one loop, there are corrections due to loops involving the squarks
and the top quark.  In the MR model,
the only \mrreborn\ one-loop correction to
the mass matrix \masss\ is a term added to the (2,2) entry (if we neglect
the bottom squark contribution which is
negligible in the allowed region of $\tan\beta$).
If the renormalization point is chosen so as to
maintain the tree level vacuum expectation of the Higgs fields, this
term takes the
form\foot{In the MSSM, there are also contributions involving
the coefficient $\mu$ of the $H_1 H_2$ mass term in the
superpotential and the coefficient $A$ of the trilinear scalar
operators in the supersymmetry
breaking terms. Those two coefficients  vanish
in the MR model because of the continuous  $U(1)_{\rm R}$ symmetry.}
\eqn\twotwomas{{\epsilon \over  \sin^2\beta} =
{3 g^2 \over 16\pi^2 \sin^2\beta}
{m_t^4 \over M_W^2}
  \ln\left({m_{\tilde t_L}^2 m_{\tilde t_R}^2 \over m_t^4}\right)\ \ .}
As discussed in \mrreborn , this correction to the Higgs masses is
quite significant; without it, LEP I would already have excluded the model
entirely.  Note that the non-degeneracy of the top squarks (Section 2)
reduces the maximum size of this correction\foot{The maximum is found
for given $m_t$ and $\tan\beta$ by
setting one top squark mass to 1 TeV
and the other to the largest value
consistent with the lower bound on $m_{\tilde\gamma}$.}
relative to the maximum size in the MSSM (achieved with degenerate
$\tilde t_L$ and $\tilde t_R$).

The dependence of the mass of the lighter CP-even neutral Higgs boson ($h$)
on $m_t$ and $\tan\beta$ is shown in \zerofigg . Note that as the top quark
gets lighter, $m_h$ becomes more restricted.  The mass of the heavier
CP-even neutral Higgs ($H$) is at least twice the size of $m_h$ throughout
the parameter space of the MR model.

In order to discuss searches for the neutral Higgs bosons of the MR model,
we need to understand the Higgs bosons' decay modes. The Higgses can
potentially decay into fermion/anti-fermion pairs (at a rate proportional
to the square of the fermion mass), chargino pairs, neutralino pairs, or
other Higgs bosons.  Due to the absence of the scalar trilinear terms in
the  soft supersymmetry breaking potential, the MR Higgs bosons do not
decay to sleptons and squarks. It is important to note that the neutral
Higgses of the MR model only couple to the neutralino pairs $\tilde z
\tilde H_\gamma$ and $\tilde z \tilde H_Z$. This is because the vanishing
of the parameters $M$, $M'$ and $\mu$ (see Section 2) in the MR model
greatly simplifies the matrix that diagonalizes the neutralino mass matrix.
We will now discuss the dominant decay modes of each neutral Higgs as a
function of its mass  \foot{In evaluating the decay widths we have included
the cosmological constraints discussed in Section 2.}.

The light CP-even Higgs boson ($h$) decays dominantly to fermion pairs,
except in the slice of parameter space where the decay $h \to A A$ is
kinematically allowed.   Therefore, the `standard' searches used to look
for $h_{SM}$ are generally suitable for $h$, with slight modifications to
take the $A A$ channel into account as must be done in searching for
$h_{MSSM}$.

In contrast, the heavy CP-even Higgs boson ($H$) is always sufficiently
massive to decay to the Higgs boson pair $h h$ or to the pair of
neutralinos  $\tilde z \tilde H_\gamma$.  Decays to fermion pairs can
dominate only when the  $H$ becomes heavy enough to open the $t \bar t$
decay channel.  As a result, the kind of searches we are going to analyze
in this paper are not suitable for $H$.

The decays of the CP-odd Higgs boson  ($A$) are shown in \firstfigg\  for
$m_t= 160$ GeV and $\tan\beta=0.71$;  the qualitative features are
independent of those particular values.  An $A$ boson lighter than roughly
the $Z$ mass will decay to the heaviest possible fermion pairs. When $m_A$
rises a bit above $m_Z$, the channel $A \to \tilde z \tilde H_\gamma$
becomes kinematically accessible and it immediately dominates.  Likewise,
the $A \to \tilde w^+ \tilde w^-$ and $A \to t\bar t$ channels each take
over once they are allowed.  The additional decay modes $A \to Z h$ and $A
\to \tilde z \tilde H_Z$ also occur for sufficiently heavy $A$.  The first
is suppressed in the MR model  by a $ZAh$ coupling factor (see Section 4)
whose value is minuscule for $A$ bosons heavy enough to decay to $Zh$.  The
second is simply never large enough to dominate.  Therefore `standard'
Higgs boson searches based on decays to fermions will be useful to probe
the region of parameter space where $m_A \lae 90$ GeV.

\newsec{`Standard' Higgs Searches at LEP I}

We now use the information we have gathered on the masses and decays of the
Higgs bosons to discuss how electron-positron collider experiments can
search for these particles.  We begin by considering the implications of
recent data from LEP experiments.

At LEP I it is possible to search for the $h$ and $A$ Higgs bosons of the
MR model; the $H$ boson is too heavy to be produced.  Specifically, the
channels
\eqn\lepi{e^+ e^- \to Z \to Z^* h, Ah}
are accessible if $h$ and $A$ are sufficiently light.  We recall  that any
$A$ boson light enough to be produced at LEP I  decays only to fermions.
The $Z Z h$ coupling is reduced relative to the standard model
$Z Z h_{SM}$ coupling
by a factor  $\sin(\beta-\alpha)$,  where $\alpha$ is the mixing
angle in the CP-even sector, determined from the diagonalization of the
mass matrix. At one loop, $\alpha$ is given by
\eqn\al{\tan 2\alpha =
{(m_A^2 + M_Z^2) \sin 2\beta \over (m_A^2 - M_Z^2) \cos 2\beta + {\epsilon
\over \sin^2\beta}},}
where $\epsilon$ was defined in eq. \twotwomas . The $Z A h$ coupling
is proportional to $\cos(\beta-\alpha)$, making the two channels
complementary.

Recent ALEPH searches for $Z^* h$ and $A h$ signals \aleph\ have produced
null results that may be translated into bounds on $\sin^2(\beta-\alpha)$
and $\cos^2(\beta-\alpha)$, respectively.  Because the $h$ mass depends on
the top quark mass (cf. equations \masss\ and \twotwomas\ ),  the bounds
also depend on $m_t$. For a relatively light top quark,
$\sin^2(\beta-\alpha)$ is almost identically 1, so that the bounds on the
MR model come exclusively from the $Z^* h$ channel.  In this case, the $h$
boson is not very heavy and the current data strongly constrain the
$\tan\beta$ vs $m_A$ plane.  Specifically, we find that {\it a top quark
lighter than 120 GeV is entirely excluded in the MR model}; for $m_t = 120$
GeV only the slice  $0.71 < \tan\beta < 0.72$ is allowed, independent of
$m_A$. The situation changes considerably as the top becomes heavier: the
constraints from  $Z^* h$ searches get weaker  (for $m_t = 200$ GeV the
constraint has evaporated) while the constraint on
$\cos^2(\beta-\alpha)$ from $A h$ production becomes stronger.

Our \secfigg\ shows the excluded regions inferred from the ALEPH  bounds on
$Z^* h$ production for several top quark masses.  The bounds from $A h$
production are only relevant for $m_t \ge 160$ GeV and tend to exclude the
small $m_A$, large $\tan\beta$ region of the plane; the bounds when $m_t =
180$ GeV and $m_t = 200$ GeV  are shown in \secfigg\ by way of
illustration.

We conclude that if the top quark is discovered to be lighter than 120 GeV
then the LEP I results just discussed will have excluded the MR model.  If
the top quark is heavier, input from future experiments will be necessary
to reach any conclusions about the MR model.

\newsec{`Standard' Searches for $Zh$ at LEP II}

The lightness of the MR model's neutral Higgs boson $h$ makes
LEP II especially useful for testing this model.  With a CM energy
$\sqrt{s} \simeq 200$ GeV, the possible
Higgs  production channels will be
\eqn\lepii{e^+ e^- \to Z^* \to Zh, Ah, AH, ZH}
The cross-section for $AH$ production is, like that for $Zh$, proportional
to $\sin^2(\beta-\alpha)$.  The complementary $ZH$ production process is
only kinematically allowed at LEP II when $m_t \leq 140 GeV$ - precisely
the region of parameter space where $\cos^2(\beta-\alpha)$ is a drastic
suppression factor.   We will ignore $H$ production since in the MR
model it does not contribute substantially to the final states for which
we propose to search.

In this section, we consider how well
LEP II will be able to search for $h$
in the $\nu\nu j j$, $l l j j$ and $j j j j$ final states arising from
production of $Zh$.  Monte Carlo results on SM Higgs boson signals and
backgrounds in these three channels
have been presented  for LEP II in ref.
\slw\ .  These show that $h_{SM}$ weighing up to 80 GeV can be detected
in $\nu\nu j j$ and $l l j j$ channels,
while one weighing up to 60 GeV can
be seen in the $j j j j$ channel with
integrated luminosity of $ 500 {\rm pb^{-1}}$ per detector.
We follow Barger et al. \barger\
in scaling from the SM simulations to estimate the
detectability of the MR model's $Z h$
signals, which differ from the SM signals
essentially by the cross-section factor
$\sin^2(\beta-\alpha)$.
To be considered `detectable' we require a signal to satisfy
\eqn\sigsee{{S \over \sqrt{B}} =
{\rm \#\ of\ signal\ events \over \sqrt{\rm \#\ of\ background\ events}}
\geq 4}
for an integrated luminosity ${\cal L}=500 {\rm pb^{-1}}$.
We add the significance ${S \over \sqrt{B}}$ of the three channels in
quadrature, considering only channels with four or more events.

As usual, the bounds depend strongly on the top quark mass.  For a light
top quark, $\sin^2(\beta-\alpha) \approx 1$
and the limits are essentially the
same as for $h_{SM}$.  When the top quark is heavier ($\gae 140$ GeV),
$\sin^2(\beta-\alpha)$ drops significantly below 1 only in the
region where $m_h$ is substantially smaller than 80 GeV; even with the
suppression factor the number of events is large. The net result is that
where $h$ decays dominantly to leptons
or jets LEP II can do as good a job of
looking for $h$ in these channels as for  $h_{SM}$

On the other hand, if a Higgs boson {\sl is} found, the fact that
$\sin^2(\beta -\alpha) \approx 1$ will make it  difficult to tell whether
$h_{SM}$ or $h$ has been seen. We have estimated the difference between the
number of  $Z h$ and $Z h_{SM}$ events, again using the simulations of  SM
Higgs boson signals presented in ref. \slw\ . We observe that the
difference is never  statistically significant (i.e. is $< 3\sigma$) when
the top is lighter than $\approx 160$ GeV.  For a heavier top, the
difference is larger than $3\sigma$ typically in the
small-$m_A$/large-$\tan\beta$ region of the parameter space (see \thirdfigg
).  However, even in the most favorable case  ($m_t = 200$ GeV) the two
models become indistinguishable when $m_A$ is heavier than 100 GeV.

It is important to note that where kinematically allowed to do so, the MR
$h$ will decay dominantly to $A A$ (except when there is an accidental
suppression of  the coupling).  In this case, the final  state topology
and, hence, the efficiency of the standard searches depends on $m_A$.
This issue has, naturally, arisen in the LEP I searches for $h_{MSSM}$.  It
was found \aleph\ that the standard searches remain effective so long as
$m_A > 2 m_b$ and slightly modified searches for `six-fermion' final states
can cover the remaining region of parameter space.  Hence our estimates
based on the scaling of the SM simulations remain reliable for $m_A > 2
m_b$.  The remaining small slice of parameter space is either  already
excluded by LEP I or can be studied using the appropriate
modified-standard searches.

Our \thirdfigg\ shows the regions in the $(\tan\beta,m_A)$  plane
accessible  to LEP II, for different values of the top quark mass. Notice
that if the top is lighter than 170 GeV,  then $h$ always weighs 80 GeV or
less.   In this case, if LEP II sees no sign of an $h$ in the standard
channels, the model's survival will depend entirely on the efficacy of the
modified-standard channels' coverage of the light $m_A$ region where $h \to
AA$ dominates.

The above estimates do not assume $b$ jet identification.
However, $b$ tagging will substantially improve
the Higgs boson discovery limit, since the $Z$-boson's
branching ratio into
$b \bar b$ is only $\sim 15\%$ while the Higgs boson decays mainly
into $b \bar b$.
It has recently been shown \delphi\ that by tagging $b$ jets
LEP II experiments will be able to detect a SM Higgs of mass
$m_{h_{SM}} \sim \sqrt{s} - 100$ GeV
\foot{We thank J.J. G\'omez-Cadenas for pointing out to us this
upgrade of the LEP detectors.}.
Within the MR model,
the regions of the $(\tan\beta,m_A)$ plane where $h$ is heavier
than 80 GeV correspond to $\sin^2(\beta -\alpha) \approx 1$,
independent of the top mass, and therefore the mass limit set for the
SM Higgs also applies for $h$. In \sixfigg\ we show the areas
of the $(\tan\beta,m_A)$ which will be probed at LEP II assuming
the $b$ tagging performances stated in ref. \delphi\ can be achieved.
The improvement is enormous.
Provided that the region where $h \to AA$ dominates can be excluded by
modified-standard searches, LEP II with a beam
energy of 95 (100) GeV will be able to rule out the MR
model if the top is lighter than 180 (200) GeV.

To conclude, standard $Zh$ searches in the MR model are strongly affected
by the fact that $\sin^2(\beta-\alpha) \approx 1$ in  most of the MR
model's parameter space. On the one hand,  a large  (i.e. visible) number
of $Z h$ events is guaranteed, so that  failure to find a neutral CP-even
Higgs boson  at LEP II would eliminate much of the $\tan\beta$ vs $m_A$
plane. On the other hand, the MR and SM Higgs bosons will look very similar
unless the top quark is quite heavy, so that further searches may well be
required to disentangle the two models.

\newsec{Searches for A and H in $\tau \tau j j$ at LEP II and beyond}

To distinguish the MR model from the SM with neutral Higgs searches,  one
must seek evidence of the $A$ and $H$ bosons in addition to the $h$.
Studies of the Higgs sector of the MSSM \slw\jangri\ suggest  exploiting
the bosons' potentially large branching fraction to tau leptons by looking
for the processes
\eqn\aaa{e^+ e^- \to Zh, Ah, ZH, AH \to \tau \tau j j.}
The scalars' masses could be deduced from the shape of the $M_{\tau \tau}$
and $M_{j j}$ invariant mass spectra of these events. In the MR model, we
have already noted that the $H$ dominantly decays to $h h$ or to $\tilde z
\tilde H_\gamma$; thus  production of $H$ bosons  will not contribute
appreciably  to $\tau \tau j j$ final states.  The $A$ boson on the other
hand, decays primarily to fermion pairs if its mass is below the $Z$ mass;
decays  to $\tau^+\tau^-$, $b\bar b$ and $c\bar c$  pairs dominate since
the scalar-fermion coupling grows  with the fermion mass.  We therefore
focus on the reactions
\eqn\bbb{e^+ e^- \to Zh, Ah \to \tau \tau j j}
in the remainder of this section.  We shall consider whether a distinct
peak due to the A boson will be directly visible and also whether slight
alterations in the $Z$ and $h$ peaks  due to the non-standard-ness of the
model would be visible.

We note that the irreducible background for the proposed signal comes from
$e^+ e^- \to Z Z$ events, which  cluster at $M_{\tau \tau}, M_{j j} = M_Z$.
As shown in \jangri\ the background would be noticeable  only when the
scalar masses are close enough to $M_Z$ for the invariant mass peaks to
overlap.

The cross-section for $A h$ production depends on the scalars' masses  and
on the suppression factor of $\cos^2(\beta - \alpha)$.  Even for the
largest attainable cross-section (essentially meaning for the smallest
value of $\sin^2(\beta - \alpha)$), we predict only about five $\tau \tau j
j$ signal events for $500 {\rm pb^{-1}}$ of integrated LEP II luminosity.
Folding in the expected 50\% detection efficiency \jangri\  makes the
signal essentially unobservable even before worrying about the precise size
of the background.  Hence LEP II does not appear capable of directly
detecting the $A$ scalar of the MR model in $\tau \tau j j$ final states.

One might also wonder whether the slight alterations in the rate of $Zh$
production in the MR model as compared with the SM would be visible in
$\tau \tau j j$. We find that, for a heavy top quark, the number of $e^+
e^- \to Zh \to \tau\tau j j$ events predicted by the two models  can differ
by of order 2$\sigma$ for small $m_A$ and large $\tan\beta$.   For example,
with $m_t$ = 200 GeV, the region with a deviation of at least 2$\sigma$ is
roughly that above the curve labeled A in \thirdfigg . When combined with
the `standard channel' signals, this could aid differentiation of the SM
and the MR model in the appropriate region of  parameter space.

Moving the search to a hypothetical NLC with a beam energy of 250 GeV and
an integrated luminosity of 10 fb$^{-1}$ changes the picture enormously.
The luminosity compensates for the  cross-section's reduction due to the
increased CM energy.  In the most favorable regions of parameter space
(small $m_A$ and large $\tan\beta$) , one might now expect of order 100
events (after including the detection efficiency); contours of number of
events are shown in \fourfigg .  This should make the $A$ boson directly
visible if $m_A \lae M_Z$.  In addition, one can more usefully exploit the
difference between  the predicted number of $Zh$ and $Zh_{SM}$ events at
the higher event rates of the NLC.  For the $\tau \tau j j$ channel, we
find that for heavy $m_t$ the significance of this difference is now quite
high:  for $m_t >$ 180 GeV (160 GeV) it is {\it less} than $3\sigma$ only
for $m_A \gae$ 100 GeV (70 GeV). Including the `standard' $Zh$ decay
channels in that comparison will naturally improve the strength of the
signal.

\newsec{Searching for Sleptons}

We have found that LEP II is not likely to be able to distinguish the MR
model from the SM by studying the Higgs sector alone.  Therefore, one
should consider other searches LEP II could make to fulfill this mission.
What immediately suggests itself is searching for superpartners.  Since
photino cosmology (Section 2) tells us that MR model sleptons weigh no more
than 140 GeV (while squarks could all have masses of a TeV), it is most
logical to search for these.

As the sleptons in the MR model are not appreciably different from those of
the MSSM (except, perhaps in their masses), we can adapt some results
obtained for the MSSM to predict how one might search for the MR sleptons.
Detailed Monte Carlo studies of slepton searches at electron-positron
machines \cdref\ have shown that sleptons are generically visible if their
masses are no more than 80\% to 90\% of the beam energy.

As \fivefigg\ shows, the slepton mass is less than 80 GeV in the MR model
so long as the top quark mass is less than about 140 GeV.  And the maximum
slepton mass for any top quark mass is about 140 GeV.   Therefore, if the
top quark is known to weigh less than 140 GeV and LEP II finds no sleptons
below a mass of 80 GeV, then the MR model will be ruled out.   Even if the
top quark mass is unknown or is greater than 140 GeV,  the allowed range of
$\tan\beta$ in the MR model will be strongly  constrained if sleptons are
not found at LEP II.  Only  an $e^+ e^-$ collider with a beam energy of at
least 175 GeV  could search the entire allowed mass range of the sleptons;
the NLC's discussed in \jangri\ meet the energy requirement easily.

It is instructive to briefly consider how useful hadron colliders are
likely to be in searching for MR model superpartners.  Judging from the
cross-sections plotted in ref \ehlq , the $4 {\rm pb^{-1}}$ of CDF's
current integrated luminosity would be expected to have produced only 4 (1)
pairs of low-rapidity 50 GeV (100 GeV) sleptons \foot{The numbers of events
given include a rapidity cut of $\vert y \vert < 1.5$ to ensure that only
`detectable' sleptons are included.}.  Hence the only sleptons light enough
to be visible at the Tevatron are already excluded by other experiments
\pdg . While squarks would be more readily produced than sleptons (due to
their color), they can also be nearly ten times as heavy; the first
consideration pales before the second.  It is only at  the higher energies
and luminosities of the SSC or LHC that the full range of either slepton or
squark masses of the MR model will be open to study.

\newsec{Discussion/Conclusions}

Because the Higgs and slepton sectors of the MR model are strongly
constrained by photino cosmology, they provide interesting search
candidates for experiments at both existing and planned electron-positron
colliders.  We have seen that studying the Higgses and sleptons provides
opportunities both for excluding the model and for distinguishing it from
the SM and the MSSM.

No matter what mass the top quark is found to have, at least one
$e^+e^-$collider will be able to use such searches to try to rule out the
MR model entirely.  Current LEP  I data on $Zh$ searches  will immediately
exclude the MR model if the top quark is found to weigh less than  120 GeV.
LEP II will have two opportunities (slepton and higgs searches)  to exclude
the model.  For $m_t < 140$ GeV, LEP II would rule out the  MR model if it
found no sleptons weighing less than 80 GeV.  For $m_t < 180$ GeV, LEP II
would rule out the model if it did not find a  neutral CP-even Higgs boson
using `standard' $Zh$ search channels. Finally, even if the top quark is as
heavy as 200 GeV,  combined searches for sleptons and neutral  Higgs bosons
at an NLC with a beam energy of at least 175 GeV  will have the potential
to exclude the MR model.

Assuming that the MR model is not directly excluded, one would naturally
wish to experimentally distinguish between it and the SM.  Two possible
approaches are (1) finding a light CP-even Higgs boson and demonstrating
that it is not $h_{SM}$ (2) finding a particle such as $A$ or $\tilde \l$
that exists in the MR model and not the SM.  We have seen that the first
approach is most useful when the top quark is very heavy.  If a neutral
CP-even Higgs boson is found at LEP II, it will be difficult to directly
tell whether $h$ or $h_{SM}$ has been found simply because
$\sin(\beta-\alpha) \approx 1$ through much of the MR parameter space.  A
deviation of the observed number of $Zh$ events from the number predicted
in the SM would only be detectable at LEP II if $m_t \gae 180$ GeV
(\thirdfigg ); the services of an NLC would be required to make this signal
useful if the top quark is lighter.   The second approach is more broadly
applicable. While the $A$ boson of the MR model will not be detectable in
$\tau \tau j j$ searches at LEP II, that collider can find sleptons
weighing up to 80 GeV. An NLC could both polish off the allowed slepton
mass range and search a respectable fraction of the parameter space in
which the $A$ decays appreciably to ordinary fermions.

One would also wish to distinguish the MR model from the MSSM.  The
difficulty of this will depend on the masses of the sleptons and Higgs
bosons.  If the sleptons are heavier than the MR model allows for given top
quark mass or if the values of $m_h$ and $m_A$ correspond to a value of
$\tan\beta$ outside the MR model range, then the choice is clear.  However
if those masses are such that either model is possible, one can still make
progress by studying the neutral Higgs bosons' decay modes.  For example,
the discovery in $\tau\tau j j$ final states of any $A$ boson at LEP II or
of an $A$ with $m_A \gae M_Z$ at an NLC would provide strong evidence for
the MSSM as opposed to the MR model.  Searching for decay modes of the
neutral Higgs bosons that are allowed in the MSSM but forbidden in the MR
model (such as decays to sleptons, squarks, and certain combinations of
neutralinos) would also be useful in disentangling the two models.

\bigbreak\bigskip\bigskip\centerline{{\bf Acknowledgements}}\nobreak
The authors thank J.J. G\'omez-Cadenas, V. Barger and L. Randall for
discussions and R.S. Chivukula for comments on the manuscript.
The authors acknowledge the hospitality of the Santa Fe Study Group where
this work was started. Research supported in part by the Texas National
Research Laboratory Commission under grant \#RGFY9206, by the National
Science Foundation under contract PHY-8714654 and by CICYT (Spain) under
Grant No. AEN90-0040. N.R. is indebted to the MEC (Spain) for a
Fulbright scholarship.

\listrefs

\listfigs

\bye